\documentclass[pra,superscriptaddress,twocolumn,floatfix]{revtex4-1}
\usepackage{graphicx}
\usepackage{dcolumn}
\usepackage{bm}
\usepackage{amsmath}
\usepackage{bbm} 
\usepackage{float}
\usepackage[english]{babel}
\usepackage[english]{babel}
\usepackage{dsfont}
\usepackage{ulem}
\usepackage{xcolor}
\graphicspath{{./img/}} 
    \renewcommand{\arraystretch}{1.2}
\newcommand{\bra}[1]{\langle #1 \vert}

\newcommand{\braOket}[3]{\langle #1 \vert #2 \vert #3 \rangle}
\newcommand{\ket}[1]{\vert #1 \rangle}

\newcommand{\mean}[1]{\langle #1 \rangle}

\begin{document}
\title{Optimizing control fields for adiabatic protocols in the presence of noise}

\author{Marllos E. F. Fernandes}
\affiliation{Departamento de F\'isica, Universidade Federal de S\~ao Carlos (UFSCar)\\ S\~ao Carlos, SP 13565-905, Brazil}
\author{Emanuel F. de Lima}
\affiliation{Departamento de F\'isica, Universidade Federal de S\~ao Carlos (UFSCar)\\ S\~ao Carlos, SP 13565-905, Brazil}


\author{Leonardo K. Castelano}
\email{lkcastelano@df.ufscar.br}
\affiliation{Departamento de F\'isica, Universidade Federal de S\~ao Carlos (UFSCar)\\ S\~ao Carlos, SP 13565-905, Brazil}
\date{\today}
\begin{abstract}
Quantum control techniques are employed to perform adiabatic quantum computing in the presence of noise. First, we analyze the adiabatic entanglement protocol (AEP) for two qubits. In this case, we found that this protocol is very robust against noise. The reason behind this fact is related to the chosen Hamiltonians, where the ground state of the initial Hamiltonian is not affected by the noise. The optimal control solution, in this case, is to leave the system in its ground state and apply a fast pulse to entangle the qubits at the end of the time evolution. Secondly, we probe a system composed of three qubits, where the goal is to teleport the first qubit to the third qubit. In this case, the ground state of the system does not share the same robustness against noise as in the case of AEP. To improve the robustness against noise, we propose the inclusion of a local control field that can drive the system to an intermediate state, which is more robust against noise in comparison to other states. The target state is also achieved by a fast pulse at the final time. We found that this approach provides a significant gain in the fidelity and can improve the adiabatic quantum computing in the so-called Noisy Intermediate-Scale Quantum (NISQ) devices in a near future.
\end{abstract}
\maketitle

\section{Introduction}

Quantum computing holds the promise of surpassing the performance of its classical counterpart, leading to a new era of technological advances. However, there exists severe bottlenecks to perform quantum computing with full power, the main one concerning the intrinsic noise of the quantum hardware, which affects the amount of quibts that can be used in practice. Nevertheless, quantum hardwares with some tens of of qubits are accessible and a great amount of effort has been devoted to operate these currently available machines, known as noisy Intermediate-Scale Quantum (NISQ) devices \cite{Preskill2018quantumcomputingin}.

Among the classes of algorithms that can run on NISQ devices are the variational quantum algorithms (VQAs) and adiabatic state preparation. VQAs are based on a properly parametrized time-dependent Hamiltonian, which is applied to a register of qubits \cite{Cerezo2021-jv,Medvidovic2021-gh}. At the end of the time evolution, the register will contain the solution for the problem.

Adiabatic quantum algorithms (AQAs) are based on a combination of time-independent Hamiltonians \cite{RevModPhys.90.015002,Barends2016}. In the adiabatic protocol, the total Hamiltonian switches adiabatically from the so-called driving-Hamiltonian, whose ground state can be in principle easily prepared, to the problem-Hamiltonian, whose ground state encodes the solution of the computational task. AQAs uses the fact that a quantum system remains in its instantaneous eigenstate given that the evolution is carried out sufficiently slowly (adiabatic theorem) \cite{doi:10.1126/science.1057726}. The drawback of this approach is the required time to keep the adiabatic theorem valid, which can be too long for real applications. Furthermore, in the presence of noise, the final ground state can be affected for sufficient long times. Several alternatives have been proposed to attack this problem, as the local adiabatic evolution or the use of counter-diabatic drivings \cite{PhysRevApplied.15.024038,PhysRevResearch.3.013227}. Another approach relies on the applications of optimal control theory (OCT) with VQA and AQA  \cite{PhysRevResearch.3.023092,PhysRevA.100.022327,PhysRevA.91.043401,Isermann2021,PhysRevA.105.032454}.
 
Recently, we investigated the problem of finding an unknown target state  using OCT in the framework of AQA, without taking into account dissipative effects~\cite{PhysRevA.105.032454}. For bounded controls, {\it i.e}, with a constraint imposed on the amplitude of the control functions, we found that the
optimal solution is obtained by simply setting both control
functions at their maximum values during the entire evolution under certain conditions. We successfully applied this solution to the adiabatic teleport protocol (ATP) considering three qubits, where the goal is to teleport the first qubit to the third qubit ~\cite{PhysRevLett.103.120504}.

  In this paper, we apply quantum control techniques to adiabatic quantum computing protocols in the presence of noise. Initially, we consider the adiabatic entanglement protocol (AEP) for two qubits. In this case, we found that this protocol is very robust against noise, meaning that it is possible to attain high fidelities for arbitrarily high values of the noise strength. The reason behind this fact is related to the chosen Hamiltonians, where the ground state of the driving-Hamiltonian is not affected by the noise. The optimal control solution, in this case, is to leave the system in its ground state and apply a fast pulse to entangle the qubits at the end of the time evolution.
  
  Subsequently, we consider the ATP, where the ground state of the system is not free from noise as in the case of AEP. To improve enhance the performance of the protocol, we propose the inclusion of a local control term in the Hamiltonian that can drive the system to an intermediate state, which is less affected by the noise in comparison to other states. As in AEP, the target state is also achieved by a fast pulse at the final time. We found that this approach provides a significant gain in the fidelity of the teleported state. This idea of Hamiltonian engineering in the context of noise suppression has been applied to NV centre in diamond~\cite{Rong2015} and in nuclear spins in quantum dots~\cite{PhysRevA.107.012601} with great success.

\section{Adiabatic entanglement protocol}

We start by considering the adiabatic protocol with two independent control functions, where are given two time-independent Hamiltonians: $H_1$ (driving) and $H_2$ (problem) \cite{PhysRevA.105.032454}. The system is initially prepared in the ground state of $H_1$, and it is desired to reach the ground state of $H_2$ at the final time. The system evolves according to the total time-dependent Hamiltonian,
\begin{align}
 H(t)=\varepsilon_1(t)H_1+\varepsilon_2(t)H_2,\label{eq2cont}
\end{align}
where $\varepsilon_1(t)$ and $\varepsilon_2(t)$ are the two independent control functions

We analyze the problem of two qubits, initially in a separable state, that should evolve to an entangled state at a given final time. In this case, the driving and the problem Hamiltonians are given by,

\begin{align}
	H_1 &= \hbar\omega_0\left(\sigma^{(1)}_z + \sigma^{(2)}_z\right) \label{entH1},\\
	H_2 &= \hbar\omega_0\left(\sigma^{(1)}_y\sigma^{(2)}_y - \sigma^{(1)}_z\sigma^{(2)}_z\right)  \label{entH2},
\end{align}
where $\sigma^{(j)}_m$ is the Pauli spin matrix in the $m$-direction acting on the $j$th-qubit, {\it e.g.}, $\sigma^{(1)}_z=\sigma_z\otimes\mathbbm{1}$. The ground states of the Hamiltonians $H_1$ and $H_2$ are given respectively by
\begin{align}
	\ket{\phi_0} &= \ket{11}, \label{eigenH1}\\
	\ket{\chi_0} &= \frac{1}{\sqrt{2}}(\ket{00} + \ket{11}), \label{eigenH2}	
\end{align}
where $\ket{\phi_0}$ is the separable initial state and $\ket{\chi_0}$ is the desired entangled state to be reached at the final time of evolution.  Under unitary dynamics (no decoherence present), the target can be achieved by the adiabatic approach with $\varepsilon_1(t)=1-\varepsilon_2(t)$ \cite{PhysRevA.105.032454}.

The non-unitary dynamics is described by the Markovian master equation for the density matrix $\rho(t)$,
\begin{equation}
\dfrac{d\rho}{dt}=-\frac{i}{\hbar}[H,\rho]+\frac{1}{2}\sum_j\gamma_j\left(2L_j\rho L_j^\dagger-L_j^\dagger L_j\rho-\rho L_j^\dagger L_j\right),\label{non-unitary}
\end{equation}
where the first term on the right-hand side represents the unitary evolution, while the second term accounts for the dissipation. $L_j$ are the Lindblad operators and $\gamma_j$ are the corresponding decay rates.

We apply Krotov method (KM) to search for optimal control fields $\varepsilon_1(t)$ and $\varepsilon_2(t)$ that maximize the fidelity concerning the final time of evolution \cite{Koch_2016,10.1063/1.3691827}. The fidelity is calculated through
\begin{equation}
    F=\left(\mathrm{Tr}\sqrt{\sqrt{\chi}\rho(T)\sqrt{\chi}}\right)^2,\label{fid}
\end{equation} 
where $\rho(T)$ is the solution of Eq.~(\ref{non-unitary}) at the final evolution time $t=T$ and $\chi$ is the corresponding density matrix of the target state. For the particular case of entanglement generation $\chi=\ket{\chi_0}\bra{\chi_0}$. We consider two different types of Lindblad operators, related to the dephasing channel and the amplitude-damping channel.
 \begin{figure}[!tb]
	\centering
	\includegraphics{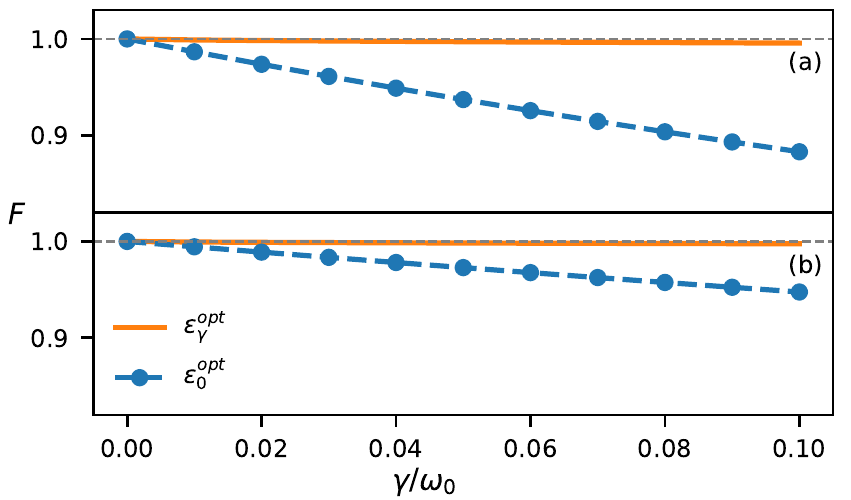}
	\caption{Fidelity evaluated by Eq.~\eqref{fid} for the AEP considering dephasing (panel (a)) and amplitude-damping (panel (b)) errors as a function of the decay rate $\gamma$ using the control function obtained from the unitary optimization (blue dotted curve) and non-unitary optimization (orange solid curve).}
	\label{ema-fxg}
\end{figure}

\begin{figure}[!tb]
	\centering
	\includegraphics{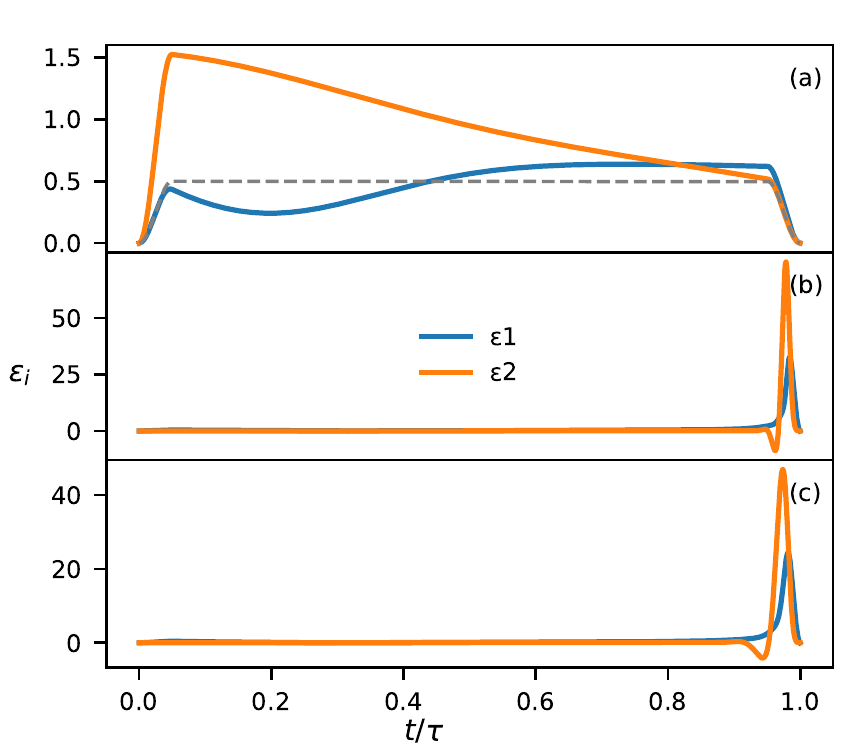}
	\caption{Optimized control functions $\varepsilon_1$ and $\varepsilon_2$ obtained with the objective of evolving the system from the state of Eq.~\eqref{eigenH1} to the entangled state of Eq.~\eqref{eigenH2}. Panel (a) refers to the unitary optimization, which does not depend on the type of noise. Panels (b) and (c) show the optimized control functions for a fixed decay rate $\gamma = 0.1$ considering the dephasing and amplitude-damping errors, respectively.}
	\label{ema-cons}
\end{figure}

\begin{figure}[!tb]
	\centering
	\includegraphics{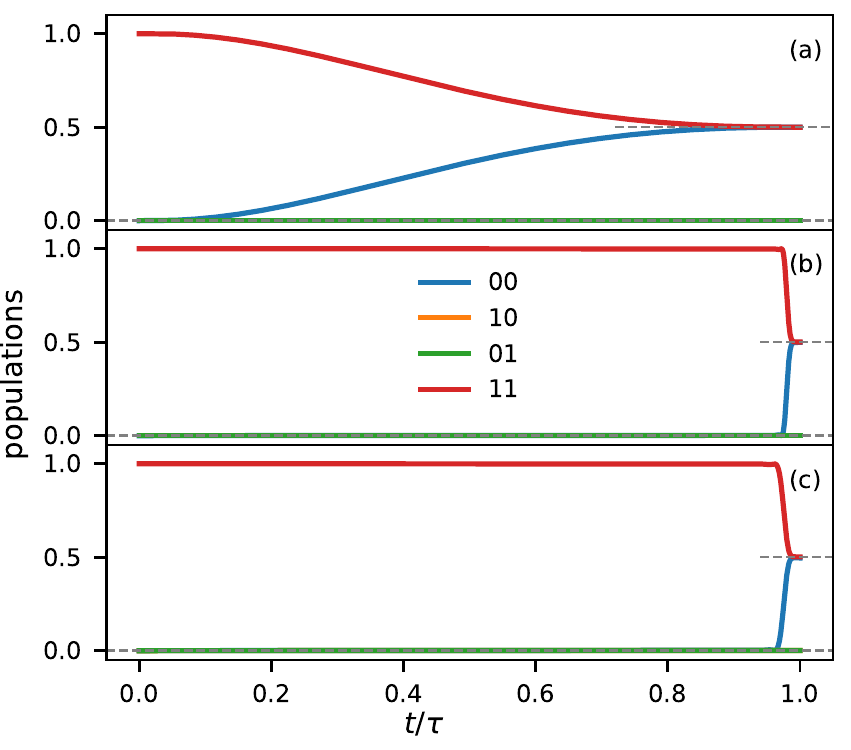}
	\caption{Population of states as a function of time for the adiabatic entanglement  protocol, resulting from the unitary evolution is shown in the panel (a). Panels (b) and (c) show the population of states as a function of time for the AEP with a fixed decay rate $\gamma = 0.1$ resulting from the dephasing and amplitude-damping noises respectively.}
	\label{ema-pops}
\end{figure}
 
 For two qubits, the sum in Eq.~(\ref{non-unitary}) contains the terms $j=1,2$ and the Lindblad operator is either $L_1=\sigma_z\otimes\mathbbm{1}$ and $L_2=\mathbbm{1}\otimes\sigma_z$ (dephasing) or $L_1=\sigma_-\otimes\mathbbm{1}$ and $L_2=\mathbbm{1}\otimes\sigma_-$ (amplitude-damping), where $\sigma_-=(\sigma_x-i\sigma_y)/2$.
 Figure~\ref{ema-fxg} shows the fidelity as a function of the decay rate $\gamma_1=\gamma_2=\gamma$ for both types of Lindblad operators, the dephasing ( panel (a) ) and the amplitude-damping  ( panel (b) ). It is noteworthy that the fidelity is always larger than $0.997$ for the considered range of values of $\gamma$ when the  optimization is evaluated using the non-unitary dynamics. Figure~\ref{ema-fxg} compares these results with the ones using unitary optimization. In this case, the control fields are obtained for the optimization carried out with $\gamma=0$, while the fidelity is evaluated through the time evolution given by Eq.~(\ref{non-unitary}) for different values of $\gamma$. For the unitary optimization, the fidelity decreases as the decay rate increases.

In order to understand the discrepancy between the unitary and the non-unitary optimization, Figure~\ref{ema-cons} shows the
 optimized controls considering the unitary ($\gamma=0$) and the non-unitary optimization ($\gamma=0.1$ for both dephasing and amplitude damping types of noise). The fields obtained from the unitary optimization (panel (a)) are smooth and have smaller amplitude than the fields obtained from the non-unitary optimization (panels (b) and (c)). Moreover, the fields resulting from the non-unitary optimization are close to zero in most of the evolution time. Only at the end of the evolution, the control fields behave as strong pulses that are able to achieve the desired target state.
 
 To further access the behavior of these controls, Fig.~\ref{ema-pops} exhibits the  population dynamics of the system. Initially only state $\ket{\psi_1}=\ket{11}$ is populated. Panel (a) presents the unitary dynamics ($\gamma=0$) with the control fields obtained from the unitary optimization. In this case, the population of the state $\ket{00}$ increases, while the population of the state $\ket{11}$ decreases as a function of time. At the final time of evolution, the population of both states $\ket{00}$ and $\ket{11}$ reaches 0.5, which is the desired value. Panels (b) and (c) show the dynamics in the presence of decoherence ($\gamma=0.1$) with the control fields obtained from the non-unitary dynamics. In contrast to panel (a), the system remains in the initially state $\ket{11}$ for as long as possible, while other states are not populated. Only at the end of the time evolution, the state $\ket{00}$ is populated and the target state is reached with very high fidelity at the final time. This result is related to the fact that the non-unitary optimization is searching for optimal fields that minimize dissipative effects. One possible way to accomplish such a task is to leave the system in states that are less affected by the dissipation. In the present case, the initial state is free from dissipation because it does not suffer from the spontaneous decay (amplitude-damping noise) or dephasing effects, because it is a separable state with no coherence. Therefore, in the dissipative case, the optimal controls are such that they leave the system essentially unperturbed up to very close to the final evolution time. Then, the controls act with a large amplitude to perform the desired transition in a short time, thus avoiding the dissipative effects.


\section{ Adiabatic teleportation protocol} 

\begin{figure}[!b]
	\centering
	\includegraphics{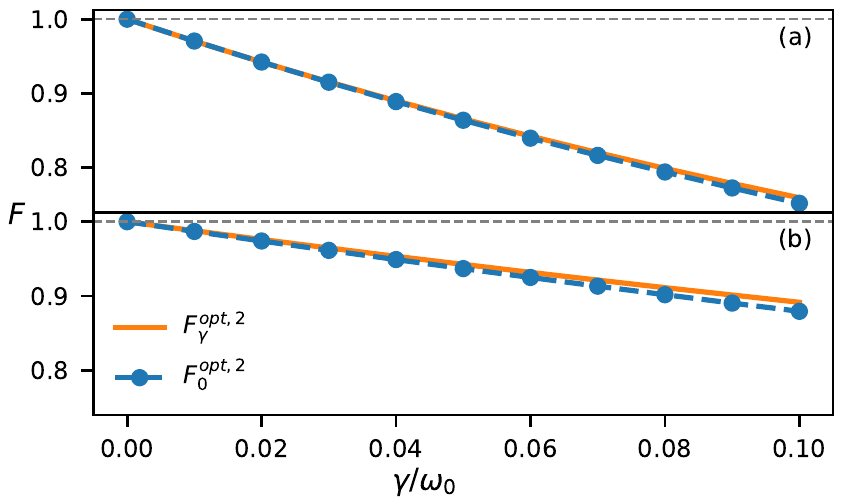}
	\caption{Fidelity evaluated by Eq.~\eqref{fid} for the teleportation protocol considering dephasing (panel (a)) and amplitude-damping (panel (b)) errors as a function of the decay rate $\gamma$. The fidelity evaluated for  two-control Hamiltonians of the type of Eq.~\eqref{eq2cont} obtained from the unitary (non-unitary) dynamics is shown by the blue dotted curve (orange solid curve).} 
	\label{tel-f1xg-2cons-cl}
\end{figure}

\begin{figure}[!t]
	\centering
	\includegraphics{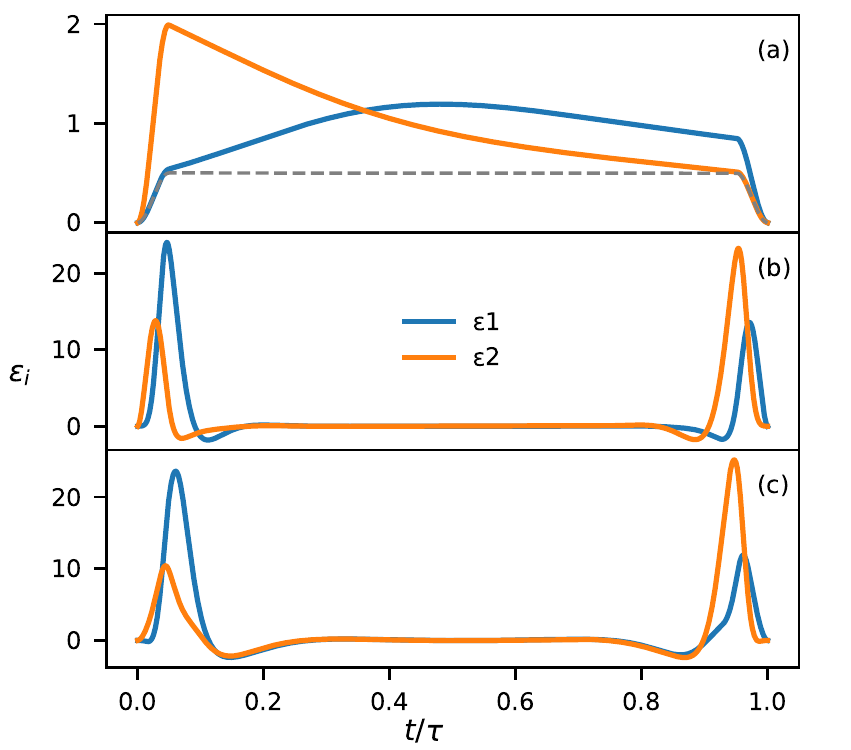}
	\caption{Optimized two-control functions $\varepsilon_i$ obtained for the adiabatic teleportation protocol using the Hamiltonian described in Eqs.~(\ref{eH0}) and (\ref{eH1}). Panel (a) refers to an unitary optimization ($\gamma = 0$). Panels (b) and (c) show the optimized two-control functions for a fixed decay rate $\gamma = 0.1$ considering the dephasing and amplitude-damping errors, respectively.}
	\label{tel-cons-2cons-cl}
\end{figure}

\begin{figure}[!tb]
	\centering
	\includegraphics{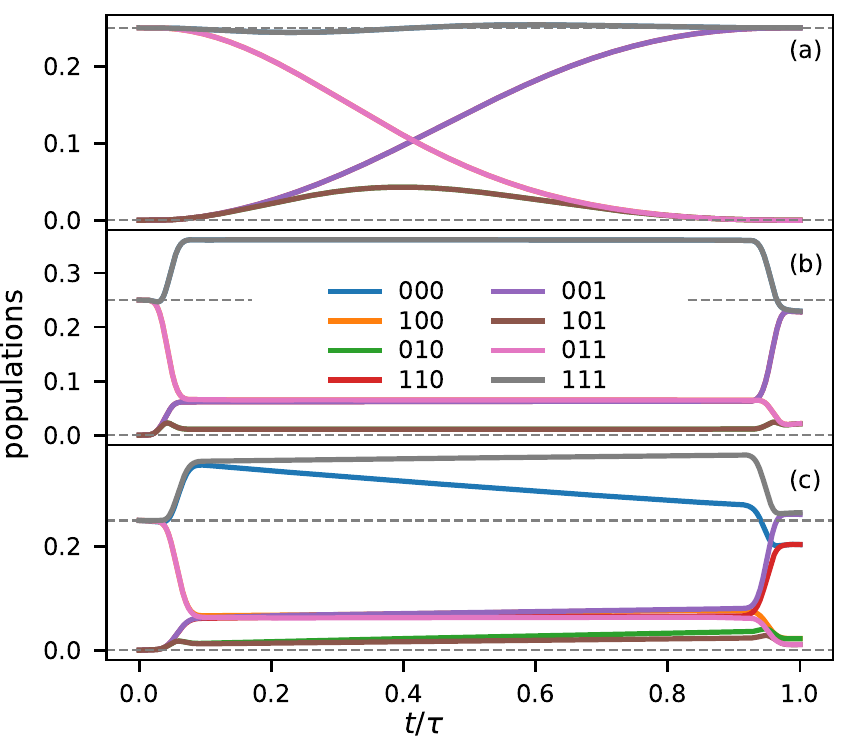}
	\caption{Population of states as a function of time for the adiabatic teleportation  protocol considering two control functions, resulting from the unitary evolution is shown in the panel (a). Panels (b) and (c) show the population of states as a function of time for the AEP with a fixed decay rate $\gamma = 0.1$ resulting from the dephasing and amplitude-damping noises respectively.}
	\label{tel-pops-2cons-cl.pdf}
\end{figure}

As a second instance of the application of OCT to the adiabatic quantum computation approach in the presence of dissipation, we consider the teleportation protocol. In this protocol, there are three qubits whose dynamics is associated to the following driving Hamiltonian,

\begin{equation}
  H_1=-\hbar\omega_0\left(\sigma^{(2)}_x\sigma^{(3)}_x+\sigma^{(2)}_z\sigma^{(3)}_z\right),\label{eH0}  
\end{equation}
for which the ground state is twofold degenerated $|\phi^{(1)}_0\rangle=|0\rangle\otimes|\Phi\rangle$ and $|\phi^{(2)}_0\rangle=|1\rangle\otimes|\Phi\rangle$, where $\ket{\Phi}$ is a Bell state, $|\Phi\rangle=\left(|00\rangle+|11\rangle\right)/\sqrt{2}$.
The problem Hamiltonian is given by

\begin{equation}
 H_2=-\hbar\omega_0\left(\sigma^{(1)}_x\sigma^{(2)}_x+\sigma^{(1)}_z\sigma^{(2)}_z\right),\label{eH1}
\end{equation}
whose ground state is also a twofold degenerated state given by $|\chi^{(1)}_0\rangle=|\Phi\rangle\otimes|0\rangle$ and $|\chi^{(2)}_0\rangle=|\Phi\rangle\otimes|1\rangle$.

This protocol aims at teleporting the information initially encoded into the first qubit to the third qubit at the final time of evolution, which is equivalent to a swap gate.
The fidelity in Eq.~(\ref{fid}) can be evaluated by considering $\ket{\chi}=
\ket{\chi_0}$, which can be chosen as any linear combination of states $\ket{\chi_0^{1}}$ and $\ket{\chi_0^{2}}$ without loss of generality~\cite{PhysRevLett.103.120504}. As already discussed in Ref.~\cite{PhysRevLett.103.120504}, one-qubit gates can be obtained by the unitary transformation of the driving-Hamiltonian $H'_1=U_GH_1U_G^\dagger$, where $U_G$ is the one-qubit gate acting on the third qubit, which can be obtained by local magnetic fields (the same idea can be generalized to implement two-qubit gates, see \cite{PhysRevLett.103.120504} for more details).

To describe dissipative effects, we choose independent Lindblad operators given by $L_i = s^{(i)}$, where $i=1,2,\textrm{and}, 3$ is the index related to the qubit that the operator is acting on, {\it e.g.}, the Lindblad operator acting on the first qubit is $L_1 = s\otimes\mathbbm{1}\otimes\mathbbm{1}$, where $s=\sigma_- = \ket{1}\bra{0}$ for amplitude-damping and $s=\sigma_z$ for dephasing.

When we apply the KM to the teleportation protocol, the resulting fidelity considering the non-unitary optimization is almost the same as the one from the unitary optimization for both dephasing and amplitude-damping, as seen in Fig.~\ref{tel-f1xg-2cons-cl}. Figure~\ref{tel-cons-2cons-cl} shows the control functions resulting from the unitary optimization (panel (a)), from non-unitary optimization with dephasing (panel (b)), and from non-unitary optimization with amplitude-damping (panel (c)). For the teleport protocol, the non-unitary optimization yields control functions that achieve amplitudes that are ten times larger than the amplitude of the unitary optimized control fields. Also, the non-unitary control fields are given by pulses at the beginning and at the end of the dynamics in contrast to the always-on form of the control fields obtained in the unitary optimization.


Figure~\ref{tel-pops-2cons-cl.pdf} shows the population dynamics corresponding to the control functions of Fig.~\ref{tel-cons-2cons-cl}. We use a linear combination of the two-fold degenerated ground state of the Hamiltonian $H_1$ as the initial state, which is given by $|\psi(0)\rangle=\frac{1}{2}\left(|0\rangle+|1\rangle\right)\otimes\left(|00\rangle+|11\rangle\right)$, while the target state is $|\psi(T)\rangle=\frac{1}{2}\left(|00\rangle+|11\rangle\right)\otimes\left(|0\rangle+|1\rangle\right)$. In the absence of noise and with the unitary-optimized control functions, the occupation of the state $\ket{011}$ smoothly decreases while increasing the amplitude of the state $\ket{001}$, also the state $\ket{101}$ is populated during the dynamics (panel (a)). In the presence of noise and with the non-unitary-optimized control functions, the occupation of the states $\ket{111}$ and $\ket{001}$ rapidly increases while decreasing the population of the state $\ket{011}$. The populations of those three states are kept constant during the dynamics until the final pulse-like portion of the controls start to act, as shown in panels (b) and (c) of Fig.~\ref{tel-cons-2cons-cl}. In this case, the occupation of the states $\ket{111}$ and $\ket{001}$ rapidly decreases, and the population of the state $\ket{011}$ increases in order to reach the target state. Although the unitary and the non-unitary optimizations use different pathways to achieve the target state, the net effect on the fidelity of the teleport protocol is essentially the same, in contrast to the AEP case. This result can be justified from the fact that the non-unitary optimization cannot avoid the effects of noise due to the symmetry of the proposed Hamiltonian.

This last result suggests that some improvement may be achieve by changing the structure of the system, for instance, by adding an extra term in the Hamiltonian. For the teleportation protocol, we propose a new Hamiltonian given by,
\begin{align}
    H=\varepsilon_1(t)H_1+\varepsilon_2(t)H_2+\varepsilon_3(t)H_3,\label{3Hs}
\end{align}
where the extra Hamiltonian is chosen as a local field $H_3=\sigma_j^{(k)}$. We have tested all combinations of local fields and obtained the results presented in Table~\ref{table_locfield}, which shows the fidelity for each type of $H_3$ for both dephasing and amplitude-damping noise considering $\gamma=0.1$. The local field that produces the highest fidelity is $H_3=\sigma_z^{(3)}$ (for dephasing) and $H_3=\sigma_z^{(2)}$ (for amplitude-damping), but local fields applied in the z-direction in any one of the qubits have a similar effect on the fidelity.

\begin{table}[!t]\label{table_locfield}
    \renewcommand{\arraystretch}{1.4}
    \centering
   \begin{tabular}{cccc}
        \hline
        H$_3$ & Dephasing &  Amplitude-damping \\
        \hline
        $\sigma_x^{(1)}$ & 0.768959 & 0.924558 \\
        $\sigma_x^{(2)}$ & 0.768192 & 0.925160 \\
        $\sigma_x^{(3)}$ & 0.770457 & 0.924571 \\
        $\sigma_y^{(1)}$ & 0.845689 & 0.961643 \\
        $\sigma_y^{(2)}$ & 0.844365 & 0.960947 \\
        $\sigma_y^{(3)}$ & 0.845620 & 0.960374 \\
        $\sigma_z^{(1)}$ & 0.898144 & 0.968651 \\
        $\sigma_z^{(2)}$ & 0.896896 & 0.970074 \\
        $\sigma_z^{(3)}$ & 0.898513 & 0.967865 \\
        \hline
    \end{tabular}
    \caption{Effects on the Fidelity of adding a local field $H_3$ in the Hamiltonian}
\end{table}

\begin{figure}[!b]
	\centering
	\includegraphics{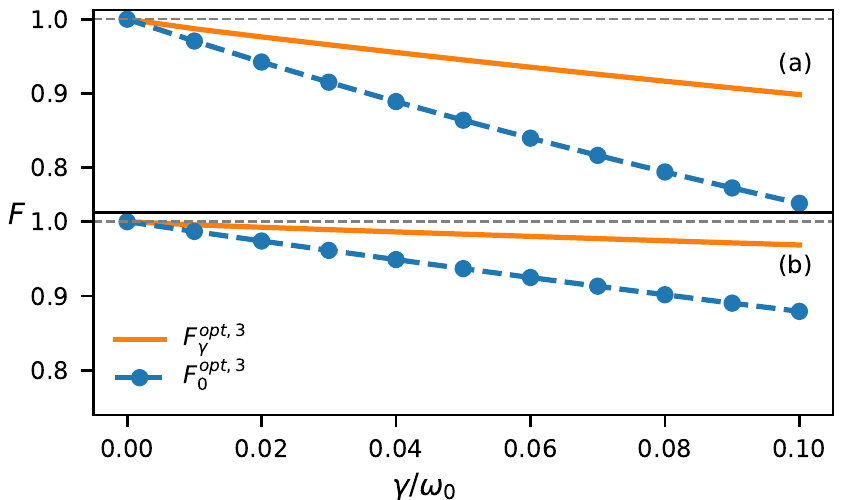}
	\caption{Fidelity evaluated by Eq.~\eqref{fid} for the teleportation protocol considering dephasing (panel (a)) and amplitude-damping (panel (b)) errors as a function of the decay rate $\gamma$. The fidelity evaluated for the optimized control functions of the Hamiltonian of Eq.~\eqref{3Hs} obtained from the unitary (non-unitary) dynamics is shown by the blue dotted curve (orange solid curve).} 
	\label{tel-f1xg-3cons-cl}
\end{figure}

 Figure~\ref{tel-f1xg-3cons-cl} shows the fidelity as a function of the decay rate for the amplitude-damping and dephasing types of noise considering the extra Hamiltonian. One can see that the fidelity evaluated with the control fields obtained from the non-unitary optimization is higher than the one calculated from the unitary optimization. For $\gamma=0.1$, the gain is 
of the order of 19\% for dephasing and of 10\% for amplitude-damping when three Hamiltonians are taken into account. This result can be understood through the analysis of the control fields and population dynamics. Again, the KM finds pulses at the beginning and at the end of the time evolution as optimal solutions for both cases, the dephasing (Fig.~8(a)) and the amplitude-damping (Fig.~8(b)) types of noise. In the dephasing case, the pulses promote a fast transition from the initial state $|\psi(0)\rangle=\frac{1}{2}\left(|0\rangle+|1\rangle\right)\otimes\left(|00\rangle+|11\rangle\right)$ to the intermediate state $|\psi_I\rangle=\frac{1}{\sqrt{2}}\left(|0\rangle+|1\rangle\right)\otimes|11\rangle$ as shown in Fig.~9 (b). Only the first qubit of this intermediate state is affected by the dephasing type of noise, which reduces the error because is only acting on one qubit. At the end of the time evolution, the pulses transform the intermediate state into the target state. In the amplitude-damping case, the intermediate state is proportional to $|\psi_I\rangle=|11\rangle\otimes\left(\alpha|0\rangle+\beta|1\rangle\right)$. Again, only the third qubit is affected by the noise as can be seen in Fig.~9 (c), where the population of the state $|110\rangle$ decreases, while the population of the state $|111\rangle$ increases as a function of time.

\begin{figure}[!t]
	\centering
	\includegraphics{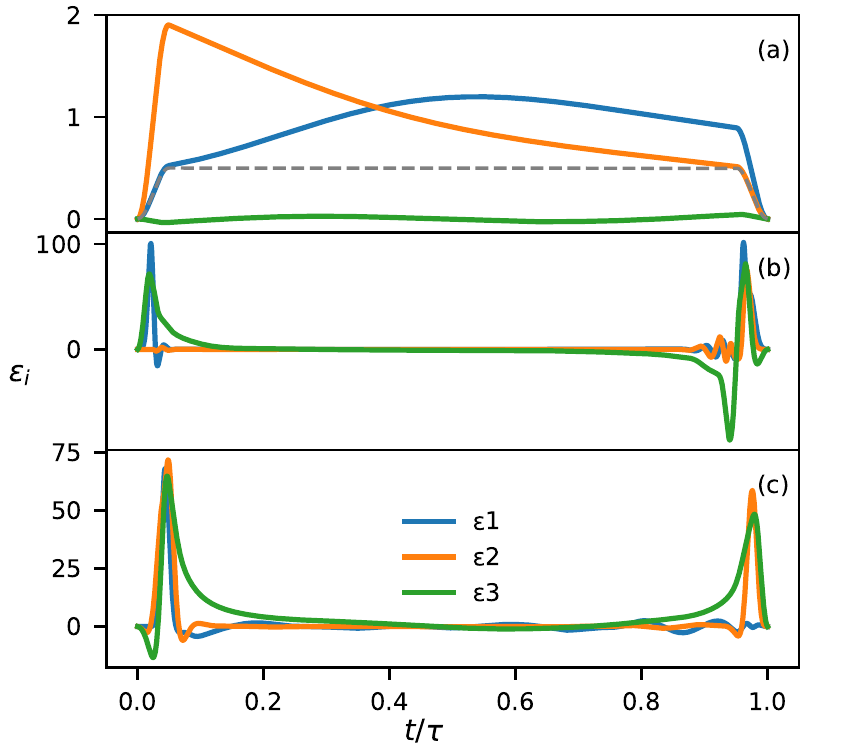}
	\caption{Optimized three-control functions $\varepsilon_i$ obtained for the adiabatic teleportation protocol using the Hamiltonian described in Eq.~(\ref{3Hs}). Panel (a) refers to an unitary optimization ($\gamma = 0$). Panels (b) and (c) show the optimized three-control functions for a fixed decay rate $\gamma = 0.1$ considering the dephasing and amplitude-damping errors, respectively.}
	\label{tel-cons-3cons-cl}
\end{figure}

\begin{figure}[!htb]
	\centering
	\includegraphics{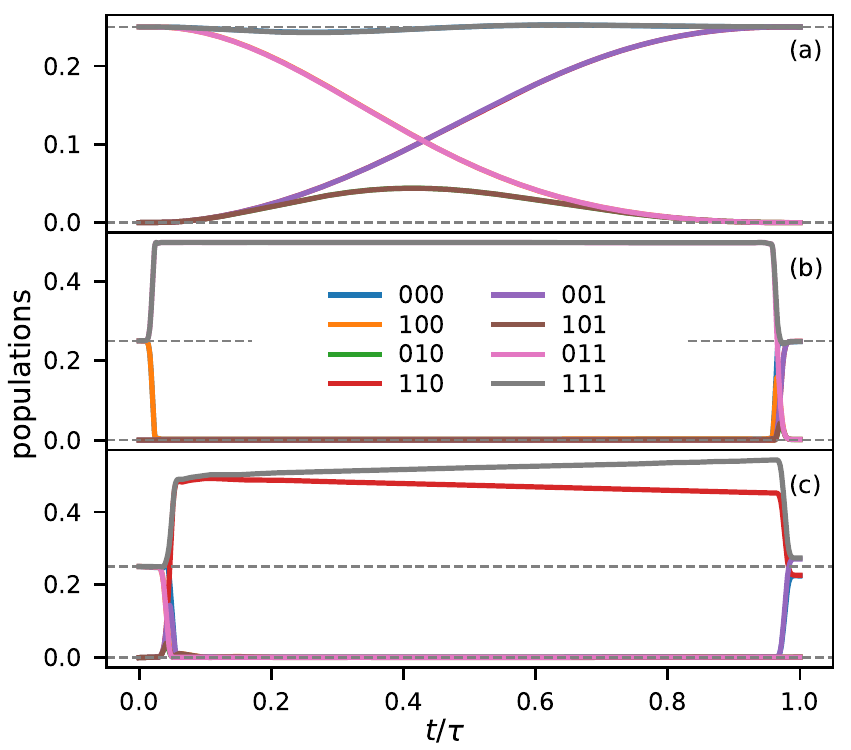}
	\caption{Population of states as a function of time for the adiabatic teleportation  protocol considering three control functions, resulting from the unitary evolution is shown in the panel (a). Panels (b) and (c) show the population of states as a function of time for the AEP with a fixed decay rate $\gamma = 0.1$ resulting from the dephasing and amplitude-damping noises respectively.}
	\label{tel-pops-3cons-cl}
\end{figure}

\begin{figure}[!tb]
	\centering
	\includegraphics{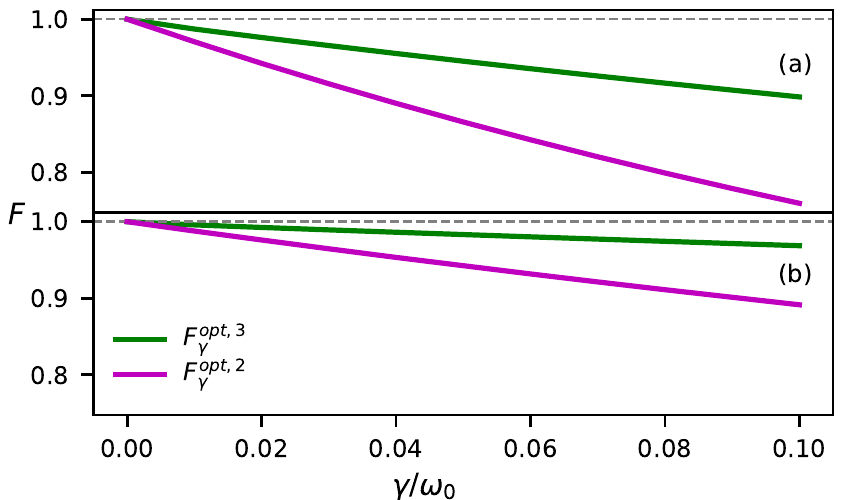}
	\caption{Mean fidelity, evaluated from Eq.~\eqref{meanfid} considering N=20000 random initial states, as a function of the decay rate $\gamma$. In panels (a) and (b), we respectively plot the results for the dephasing and the amplitude-damping types of error considering the mean fidelity obtained for two- $\mean{F_\gamma^2}$ and three-control $\mean{F_\gamma^3}$ functions.} 
	\label{tel-f1xg2-3cons-cl}
\end{figure}

Such results show that an appropriate inclusion of an extra local Hamiltonian can improve the fidelity because different pathways can be accessed. As a proof of the robustness of the inclusion of the extra Hamiltonian approach, we evaluate the mean fidelity considering a myriad of different states that must be teleported from the first to the third qubit. The mean fidelity is evaluated according to the following formula,
\begin{equation}
	\mean{F} = \frac{1}{N} \sum_{i=1}^{N} \braOket{\psi_i}{O^\dagger\rho_i(T)O}{\psi_i},\label{meanfid}
\end{equation}
where $\ket{\psi_i}=(\alpha\ket{0}+\beta\ket{1})\otimes\ket{\Phi}$ is a particular initial state from a set of N states, $O\ket{\psi_i}=\ket{\Phi}\otimes(\alpha\ket{0}+\beta\ket{1})$, and $\rho_i(T)$ is the density matrix corresponding to the time evolution of the initial state $\ket{\psi_i}\bra{\psi_i}$ by the Eq.~(\ref{non-unitary}). The complex numbers $\alpha$ and $\beta$ are random numbers generated through the normal distribution ~\cite{Sommers_2004}. In Fig.~10, we plot the mean fidelity as a function of the decay rate for N=20000 states, considering the optimized fields evaluated only for the initial state $|\psi(0)\rangle=\frac{1}{2}\left(|0\rangle+|1\rangle\right)\otimes\left(|00\rangle+|11\rangle\right)$. The mean fidelity denoted by $\mean{F_\gamma^3}$ and $\mean{F_\gamma^2}$ is evaluated considering the total Hamiltonian with or without the extra term, respectively. One can see that the mean fidelity evaluated considering the extra term $\mean{F_\gamma^3}$ has a better performance than the $\mean{F_\gamma^2}$. For example, the gain of including the extra term in the mean fidelity is of the order of 18\% (see Fig.~10(a)) for dephasing and 8\% (see Fig.~10(b)) for amplitude-damping when $\gamma=0.1$.

\section{Conclusion}

In this work we apply the KM for open quantum systems (non-unitary optimization) to numerically investigate the performance of adiabatic quantum computing. We considered two different protocols aimed at performing entanglement and teleportation. In the AEP, we verified that the non-unitary optimization performs much better compared to the unitary optimization. The reason for the enhancement of the fidelity is related to the finding of pathways that avoid noise effects. The strategy used by the optimized controls is essentially to leave the system in the states that are not affected by the noise, and only close to the final time, to act abruptly for driving the system to the desired state, thus spending only a very short time in the states that suffer from the noise effects. In contrast, the unitary optimization, not aware of noise effects, drives the system to the affected states, not being capable of avoiding the dissipation.

We applied the same analysis for the teleport protocol. However, in this case, the fidelity obtained from the non-unitary optimization is similar to the obtained from the unitary optimization. We conjecture that the reason for such comparatively low performance of the non-unitary optimization is related to the type of the Hamiltonian used in the adiabatic teleportation protocol. The proposed Hamiltonian does not allow for pathways that are free from noise effects. To circumvent this problem, we propose the inclusion of an extra local Hamiltonian. We found that the optimal control function derived by the non-unitary optimization through the inclusion of this extra term performs better in comparison to the unitary optimization. The gain in the mean fidelity can be larger than 8\% for the highest value of decay rate considered in our calculations. We believe that the combination of OCT for open quantum systems with extra local terms in the original protocols can be very interesting in real applications of AQC. Since this conclusion has been reached from a set of limited cases, additional investigation has to be carried out to verify its extension to other cases.

\begin{acknowledgements}
The authors are grateful for financial support by  the Brazilian Agencies FAPESP, CNPq and CAPES. 
LKC and EFL thanks to the Brazilian Agencies FAPESP (grants 2019/09624-3 and 2014/23648-9) and CNPq (grants 311450/2019-9 and 423982/2018-4 ) for supporting this research. 
\end{acknowledgements}


\begin{thebibliography}{19}
	\expandafter\ifx\csname natexlab\endcsname\relax\def\natexlab#1{#1}\fi
	\expandafter\ifx\csname bibnamefont\endcsname\relax
	\def\bibnamefont#1{#1}\fi
	\expandafter\ifx\csname bibfnamefont\endcsname\relax
	\def\bibfnamefont#1{#1}\fi
	\expandafter\ifx\csname citenamefont\endcsname\relax
	\def\citenamefont#1{#1}\fi
	\expandafter\ifx\csname url\endcsname\relax
	\def\url#1{\texttt{#1}}\fi
	\expandafter\ifx\csname urlprefix\endcsname\relax\def\urlprefix{URL }\fi
	\providecommand{\bibinfo}[2]{#2}
	\providecommand{\eprint}[2][]{\url{#2}}
	
	\bibitem[{\citenamefont{Preskill}(2018)}]{Preskill2018quantumcomputingin}
	\bibinfo{author}{\bibfnamefont{J.}~\bibnamefont{Preskill}},
	\bibinfo{journal}{{Quantum}} \textbf{\bibinfo{volume}{2}},
	\bibinfo{pages}{79} (\bibinfo{year}{2018}).
	
	\bibitem[{\citenamefont{Cerezo et~al.}(2021)\citenamefont{Cerezo, Arrasmith,
			Babbush, Benjamin, Endo, Fujii, McClean, Mitarai, Yuan, Cincio
			et~al.}}]{Cerezo2021-jv}
	\bibinfo{author}{\bibfnamefont{M.}~\bibnamefont{Cerezo}},
	\bibinfo{author}{\bibfnamefont{A.}~\bibnamefont{Arrasmith}},
	\bibinfo{author}{\bibfnamefont{R.}~\bibnamefont{Babbush}},
	\bibinfo{author}{\bibfnamefont{S.~C.} \bibnamefont{Benjamin}},
	\bibinfo{author}{\bibfnamefont{S.}~\bibnamefont{Endo}},
	\bibinfo{author}{\bibfnamefont{K.}~\bibnamefont{Fujii}},
	\bibinfo{author}{\bibfnamefont{J.~R.} \bibnamefont{McClean}},
	\bibinfo{author}{\bibfnamefont{K.}~\bibnamefont{Mitarai}},
	\bibinfo{author}{\bibfnamefont{X.}~\bibnamefont{Yuan}},
	\bibinfo{author}{\bibfnamefont{L.}~\bibnamefont{Cincio}},
	\bibnamefont{et~al.}, \bibinfo{journal}{Nature Reviews Physics}
	\textbf{\bibinfo{volume}{3}}, \bibinfo{pages}{625} (\bibinfo{year}{2021}).
	
	\bibitem[{\citenamefont{Medvidovi{\'c} and Carleo}(2021)}]{Medvidovic2021-gh}
	\bibinfo{author}{\bibfnamefont{M.}~\bibnamefont{Medvidovi{\'c}}}
	\bibnamefont{and} \bibinfo{author}{\bibfnamefont{G.}~\bibnamefont{Carleo}},
	\bibinfo{journal}{npj Quantum Information} \textbf{\bibinfo{volume}{7}},
	\bibinfo{pages}{101} (\bibinfo{year}{2021}).
	
	\bibitem[{\citenamefont{Albash and Lidar}(2018)}]{RevModPhys.90.015002}
	\bibinfo{author}{\bibfnamefont{T.}~\bibnamefont{Albash}} \bibnamefont{and}
	\bibinfo{author}{\bibfnamefont{D.~A.} \bibnamefont{Lidar}},
	\bibinfo{journal}{Rev. Mod. Phys.} \textbf{\bibinfo{volume}{90}},
	\bibinfo{pages}{015002} (\bibinfo{year}{2018}).
	
	\bibitem[{\citenamefont{Barends et~al.}(2016)\citenamefont{Barends, Shabani,
			Lamata, Kelly, Mezzacapo, Heras, Babbush, Fowler, Campbell, Chen
			et~al.}}]{Barends2016}
	\bibinfo{author}{\bibfnamefont{R.}~\bibnamefont{Barends}},
	\bibinfo{author}{\bibfnamefont{A.}~\bibnamefont{Shabani}},
	\bibinfo{author}{\bibfnamefont{L.}~\bibnamefont{Lamata}},
	\bibinfo{author}{\bibfnamefont{J.}~\bibnamefont{Kelly}},
	\bibinfo{author}{\bibfnamefont{A.}~\bibnamefont{Mezzacapo}},
	\bibinfo{author}{\bibfnamefont{U.~L.} \bibnamefont{Heras}},
	\bibinfo{author}{\bibfnamefont{R.}~\bibnamefont{Babbush}},
	\bibinfo{author}{\bibfnamefont{A.~G.} \bibnamefont{Fowler}},
	\bibinfo{author}{\bibfnamefont{B.}~\bibnamefont{Campbell}},
	\bibinfo{author}{\bibfnamefont{Y.}~\bibnamefont{Chen}}, \bibnamefont{et~al.},
	\bibinfo{journal}{Nature} \textbf{\bibinfo{volume}{534}},
	\bibinfo{pages}{222} (\bibinfo{year}{2016}).
	
	\bibitem[{\citenamefont{Farhi et~al.}(2001)\citenamefont{Farhi, Goldstone,
			Gutmann, Lapan, Lundgren, and Preda}}]{doi:10.1126/science.1057726}
	\bibinfo{author}{\bibfnamefont{E.}~\bibnamefont{Farhi}},
	\bibinfo{author}{\bibfnamefont{J.}~\bibnamefont{Goldstone}},
	\bibinfo{author}{\bibfnamefont{S.}~\bibnamefont{Gutmann}},
	\bibinfo{author}{\bibfnamefont{J.}~\bibnamefont{Lapan}},
	\bibinfo{author}{\bibfnamefont{A.}~\bibnamefont{Lundgren}}, \bibnamefont{and}
	\bibinfo{author}{\bibfnamefont{D.}~\bibnamefont{Preda}},
	\bibinfo{journal}{Science} \textbf{\bibinfo{volume}{292}},
	\bibinfo{pages}{472} (\bibinfo{year}{2001}).
	
	\bibitem[{\citenamefont{Hegade et~al.}(2021)\citenamefont{Hegade, Paul, Ding,
			Sanz, Albarr\'an-Arriagada, Solano, and Chen}}]{PhysRevApplied.15.024038}
	\bibinfo{author}{\bibfnamefont{N.~N.} \bibnamefont{Hegade}},
	\bibinfo{author}{\bibfnamefont{K.}~\bibnamefont{Paul}},
	\bibinfo{author}{\bibfnamefont{Y.}~\bibnamefont{Ding}},
	\bibinfo{author}{\bibfnamefont{M.}~\bibnamefont{Sanz}},
	\bibinfo{author}{\bibfnamefont{F.}~\bibnamefont{Albarr\'an-Arriagada}},
	\bibinfo{author}{\bibfnamefont{E.}~\bibnamefont{Solano}}, \bibnamefont{and}
	\bibinfo{author}{\bibfnamefont{X.}~\bibnamefont{Chen}},
	\bibinfo{journal}{Phys. Rev. Applied} \textbf{\bibinfo{volume}{15}},
	\bibinfo{pages}{024038} (\bibinfo{year}{2021}).
	
	\bibitem[{\citenamefont{Prielinger et~al.}(2021)\citenamefont{Prielinger,
			Hartmann, Yamashiro, Nishimura, Lechner, and
			Nishimori}}]{PhysRevResearch.3.013227}
	\bibinfo{author}{\bibfnamefont{L.}~\bibnamefont{Prielinger}},
	\bibinfo{author}{\bibfnamefont{A.}~\bibnamefont{Hartmann}},
	\bibinfo{author}{\bibfnamefont{Y.}~\bibnamefont{Yamashiro}},
	\bibinfo{author}{\bibfnamefont{K.}~\bibnamefont{Nishimura}},
	\bibinfo{author}{\bibfnamefont{W.}~\bibnamefont{Lechner}}, \bibnamefont{and}
	\bibinfo{author}{\bibfnamefont{H.}~\bibnamefont{Nishimori}},
	\bibinfo{journal}{Phys. Rev. Research} \textbf{\bibinfo{volume}{3}},
	\bibinfo{pages}{013227} (\bibinfo{year}{2021}).
	
	\bibitem[{\citenamefont{Choquette et~al.}(2021)\citenamefont{Choquette,
			Di~Paolo, Barkoutsos, S\'en\'echal, Tavernelli, and
			Blais}}]{PhysRevResearch.3.023092}
	\bibinfo{author}{\bibfnamefont{A.}~\bibnamefont{Choquette}},
	\bibinfo{author}{\bibfnamefont{A.}~\bibnamefont{Di~Paolo}},
	\bibinfo{author}{\bibfnamefont{P.~K.} \bibnamefont{Barkoutsos}},
	\bibinfo{author}{\bibfnamefont{D.}~\bibnamefont{S\'en\'echal}},
	\bibinfo{author}{\bibfnamefont{I.}~\bibnamefont{Tavernelli}},
	\bibnamefont{and} \bibinfo{author}{\bibfnamefont{A.}~\bibnamefont{Blais}},
	\bibinfo{journal}{Phys. Rev. Research} \textbf{\bibinfo{volume}{3}},
	\bibinfo{pages}{023092} (\bibinfo{year}{2021}).
	
	\bibitem[{\citenamefont{Lin et~al.}(2019)\citenamefont{Lin, Wang, Kolesov, and
			Kalabi\ifmmode~\acute{c}\else \'{c}\fi{}}}]{PhysRevA.100.022327}
	\bibinfo{author}{\bibfnamefont{C.}~\bibnamefont{Lin}},
	\bibinfo{author}{\bibfnamefont{Y.}~\bibnamefont{Wang}},
	\bibinfo{author}{\bibfnamefont{G.}~\bibnamefont{Kolesov}}, \bibnamefont{and}
	\bibinfo{author}{\bibfnamefont{U.~c.~v.}
		\bibnamefont{Kalabi\ifmmode~\acute{c}\else \'{c}\fi{}}},
	\bibinfo{journal}{Phys. Rev. A} \textbf{\bibinfo{volume}{100}},
	\bibinfo{pages}{022327} (\bibinfo{year}{2019}).
	
	\bibitem[{\citenamefont{Riviello et~al.}(2015)\citenamefont{Riviello, Tibbetts,
			Brif, Long, Wu, Ho, and Rabitz}}]{PhysRevA.91.043401}
	\bibinfo{author}{\bibfnamefont{G.}~\bibnamefont{Riviello}},
	\bibinfo{author}{\bibfnamefont{K.~M.} \bibnamefont{Tibbetts}},
	\bibinfo{author}{\bibfnamefont{C.}~\bibnamefont{Brif}},
	\bibinfo{author}{\bibfnamefont{R.}~\bibnamefont{Long}},
	\bibinfo{author}{\bibfnamefont{R.-B.} \bibnamefont{Wu}},
	\bibinfo{author}{\bibfnamefont{T.-S.} \bibnamefont{Ho}}, \bibnamefont{and}
	\bibinfo{author}{\bibfnamefont{H.}~\bibnamefont{Rabitz}},
	\bibinfo{journal}{Phys. Rev. A} \textbf{\bibinfo{volume}{91}},
	\bibinfo{pages}{043401} (\bibinfo{year}{2015}).
	
	\bibitem[{\citenamefont{Isermann}(2021)}]{Isermann2021}
	\bibinfo{author}{\bibfnamefont{S.}~\bibnamefont{Isermann}},
	\bibinfo{journal}{Quantum Information Processing}
	\textbf{\bibinfo{volume}{20}}, \bibinfo{pages}{300} (\bibinfo{year}{2021}).
	
	\bibitem[{\citenamefont{de~Lima et~al.}(2022)\citenamefont{de~Lima, Fernandes,
			and Castelano}}]{PhysRevA.105.032454}
	\bibinfo{author}{\bibfnamefont{E.~F.} \bibnamefont{de~Lima}},
	\bibinfo{author}{\bibfnamefont{M.~E.~F.} \bibnamefont{Fernandes}},
	\bibnamefont{and} \bibinfo{author}{\bibfnamefont{L.~K.}
		\bibnamefont{Castelano}}, \bibinfo{journal}{Phys. Rev. A}
	\textbf{\bibinfo{volume}{105}}, \bibinfo{pages}{032454}
	(\bibinfo{year}{2022}).
	
	\bibitem[{\citenamefont{Bacon and Flammia}(2009)}]{PhysRevLett.103.120504}
	\bibinfo{author}{\bibfnamefont{D.}~\bibnamefont{Bacon}} \bibnamefont{and}
	\bibinfo{author}{\bibfnamefont{S.~T.} \bibnamefont{Flammia}},
	\bibinfo{journal}{Phys. Rev. Lett.} \textbf{\bibinfo{volume}{103}},
	\bibinfo{pages}{120504} (\bibinfo{year}{2009}).
	
	\bibitem[{\citenamefont{Rong et~al.}(2015)\citenamefont{Rong, Geng, Shi, Liu,
			Xu, Ma, Kong, Jiang, Wu, and Du}}]{Rong2015}
	\bibinfo{author}{\bibfnamefont{X.}~\bibnamefont{Rong}},
	\bibinfo{author}{\bibfnamefont{J.}~\bibnamefont{Geng}},
	\bibinfo{author}{\bibfnamefont{F.}~\bibnamefont{Shi}},
	\bibinfo{author}{\bibfnamefont{Y.}~\bibnamefont{Liu}},
	\bibinfo{author}{\bibfnamefont{K.}~\bibnamefont{Xu}},
	\bibinfo{author}{\bibfnamefont{W.}~\bibnamefont{Ma}},
	\bibinfo{author}{\bibfnamefont{F.}~\bibnamefont{Kong}},
	\bibinfo{author}{\bibfnamefont{Z.}~\bibnamefont{Jiang}},
	\bibinfo{author}{\bibfnamefont{Y.}~\bibnamefont{Wu}}, \bibnamefont{and}
	\bibinfo{author}{\bibfnamefont{J.}~\bibnamefont{Du}},
	\bibinfo{journal}{Nature Communications} \textbf{\bibinfo{volume}{6}},
	\bibinfo{pages}{8748} (\bibinfo{year}{2015}).
	
	\bibitem[{\citenamefont{Jing et~al.}(2023)\citenamefont{Jing, Du, Tang, and
			Zhang}}]{PhysRevA.107.012601}
	\bibinfo{author}{\bibfnamefont{L.}~\bibnamefont{Jing}},
	\bibinfo{author}{\bibfnamefont{P.}~\bibnamefont{Du}},
	\bibinfo{author}{\bibfnamefont{H.}~\bibnamefont{Tang}}, \bibnamefont{and}
	\bibinfo{author}{\bibfnamefont{W.}~\bibnamefont{Zhang}},
	\bibinfo{journal}{Phys. Rev. A} \textbf{\bibinfo{volume}{107}},
	\bibinfo{pages}{012601} (\bibinfo{year}{2023}).
	
	\bibitem[{\citenamefont{Koch}(2016)}]{Koch_2016}
	\bibinfo{author}{\bibfnamefont{C.~P.} \bibnamefont{Koch}},
	\bibinfo{journal}{Journal of Physics: Condensed Matter}
	\textbf{\bibinfo{volume}{28}}, \bibinfo{pages}{213001}
	(\bibinfo{year}{2016}).
	
	\bibitem[{\citenamefont{Reich et~al.}(2012)\citenamefont{Reich, Ndong, and
			Koch}}]{10.1063/1.3691827}
	\bibinfo{author}{\bibfnamefont{D.~M.} \bibnamefont{Reich}},
	\bibinfo{author}{\bibfnamefont{M.}~\bibnamefont{Ndong}}, \bibnamefont{and}
	\bibinfo{author}{\bibfnamefont{C.~P.} \bibnamefont{Koch}},
	\bibinfo{journal}{The Journal of Chemical Physics}
	\textbf{\bibinfo{volume}{136}} (\bibinfo{year}{2012}),
	\bibinfo{note}{104103}.
	
	\bibitem[{\citenamefont{Sommers and Zyczkowski}(2004)}]{Sommers_2004}
	\bibinfo{author}{\bibfnamefont{H.-J.} \bibnamefont{Sommers}} \bibnamefont{and}
	\bibinfo{author}{\bibfnamefont{K.}~\bibnamefont{Zyczkowski}},
	\bibinfo{journal}{Journal of Physics A: Mathematical and General}
	\textbf{\bibinfo{volume}{37}}, \bibinfo{pages}{8457} (\bibinfo{year}{2004}).
	
\end{thebibliography}
\end{document}